\documentclass[twocolumn]{svjour3}       
\smartqed  
\usepackage{graphicx}
\usepackage{cite}
\usepackage{amsmath,amssymb,amsfonts}
\usepackage{algorithmic}
\usepackage{textcomp}
\usepackage{xcolor}

\begin{document}

\title{Machine Learning Based Anxiety Detection in Older Adults using Wristband Sensors and Context Feature\\}


\author{Rajdeep Kumar Nath         \and
        Himanshu Thapliyal* 
}


\institute{Rajdeep Kumar Nath \at
              University of Kentucky \\
              Lexington, Kentucky, USA\\         
           \and
           *Himanshu Thapliyal \at
              University of Kentucky \\
              Lexington, Kentucky, USA\\
              Corresponding Author: hthapliyal@uky.edu
}

\date{Received: date / Accepted: date}

\maketitle

\begin{abstract}
This paper explores a novel method for anxiety detection in older adults using simple wristband sensors such as Electrodermal Activity (EDA) and Photoplethysmogram (PPG) and a context-based feature. The proposed method for anxiety detection combines features from a single physiological signal with an experimental context-based feature to improve the performance of the anxiety detection model. The experimental data for this work is obtained from a year-long experiment on 41 healthy older adults (26 females and 15 males) in the age range
60-80 with mean age 73.36 ± 5.25 during a Trier
Social Stress Test (TSST) protocol. The anxiety level ground truth was obtained from State-Trait Anxiety Inventory (STAI), which is regarded as the gold standard to measure perceived anxiety. EDA and Blood Volume Pulse (BVP) signals were recorded using a wrist-worn EDA and PPG sensor respectively. 47 features were computed from EDA and BVP signal, out of which a final set of 24 significantly correlated features were selected for analysis. The phases of the experimental study are encoded as unique integers to generate the context feature vector. A combination of features from a single sensor with the context feature vector is used for training a machine learning model to distinguish between anxious and not-anxious states. Results and analysis showed that the EDA and BVP machine learning models that combined the context feature along with the physiological features achieved 3.37\% and 6.41\% higher accuracy respectively than the models that used only physiological features. Further, end-to-end processing of EDA and BVP signals was simulated for real-time anxiety level detection. This work demonstrates the practicality of the proposed anxiety detection method in facilitating long-term monitoring of anxiety in older adults using low-cost consumer devices.  
\end{abstract}


\section{Introduction}
Anxiety is a complex emotional and behavioral response that results from the anticipation of a negative situation (real or perceived) that is potentially perceived as harmful to the individual \cite{001,002}. Like stress, repeated exposure to anxiety can cause psychological disorders such as clinical depression and anxiety disorder \cite{003}. Further, chronic anxiety has also been found to be strongly associated with physiological abnormalities such as cardiovascular diseases, insomnia, and impaired cognitive ability \cite{004,006,005}. Although anxiety is a global psychological disorder affecting all population groups, a qualitative difference has been observed between the anxiety experienced by an older adult to that of a younger adult \cite{007}. Statistics show that about 15\% to 20\% of older adults suffer from depression \cite{008} and 15\% to 40\% of older adults suffer from anxiety \cite{009}. The increased perception of anxiety and its consequent effects in older adults is attributed to several factors prevalent among older adults such as loneliness from retirement, decreased physical abilities, etc. \cite{010,011}.  Hence, monitoring anxiety on a regular basis to ensure proper management of anxiety is important in reducing the resulting harmful and irreversible effects of anxiety in older adults \cite{012}. 

Typically, anxiety is detected using a self-reported questionnaire \cite{013,014}. However, such diagnostic methods lack objectivity, are influenced by subjective bias and do not facilitate continuous unobtrusive monitoring of anxiety during activities of daily living \cite{016,015}. Recently, significant research interest has been drawn towards using features from physiological signals to train a machine learning model for detecting anxiety \cite{016,017,018,015}. This method has the potential to automate the clinical diagnosis of anxiety-related disorder as minimum human intervention is required. Further, integrating frameworks for anxiety detection by monitoring physiological signals with wearable devices can support continuous monitoring of anxiety levels. However, most of the methods explored in the literature focused mostly on ECG (Electrocardiogram), and EEG (Electroencephalograph) signals \cite{018,015}. Researchers have also explored fusing features from other physiological signals such as PPG (Photoplethysmography) and eye movement \cite{019,015}. However, EEG and ECG signals have complex acquisition setup and may not be suitable for developing low-power and low-cost anxiety detection systems for consumer end devices. Moreover, although fusing different physiological features increases the performance of the detection model, the complexity of integrating and processing multiple physiological signals in real-time could be a significant design challenge. 

In this paper, we propose a machine learning based anxiety detection system for older adults. Our anxiety detection system fuses an experimental context-based feature with physiological features from a single physiological signal in improving the performance of the anxiety detection model for older adults. We have evaluated the performance of two physiological signals, EDA (Electrodermal Activity), and BVP (Blood Volume Pulse), in conjunction with an experimental context feature to train a machine learning model to distinguish between anxious and non-anxious states. 

The novel contribution of the proposed work is listed as follows:

\begin{itemize}
\item A machine learning based anxiety detection system is proposed for detecting anxiety in older adults. 
\item Data of 41 healthy older adults were collected and analyzed for developing, validating, and testing the proposed anxiety detection system.
\item Features from two physiological signals, EDA and BVP along with a context-based feature was used in training a machine learning classifier to distinguish between anxious and not-anxious states.
\item Extensive analysis is performed to highlight the importance of the context-based feature in reducing misclassification of the anxiety detection model.  
\item The feasibility of the proposed system in detecting anxiety in real-time is analyzed by simulating end-to-end anxiety level classification task. 
\end{itemize}

The paper is organized as follows, section \ref{sec_related_work} will discuss the related work in the context of using wearable sensors and machine learning techniques for detecting anxiety. Section \ref{sec_proposed_method} will present the proposed method for anxiety detection. Section \ref{sec_experiment} will present the experimental setup adopted for the study. Section \ref{sec_results} will present the results and analysis and finally section \ref{sec_conclusion} will conclude the article. 

\section{Related Work}
\label{sec_related_work}
Unlike stress detection, anxiety detection is relatively less explored in the context of the Internet of Medical Things (IoMT). For example, there is a significant amount of research work available in the literature that uses low-cost consumer electronic components to monitor stress continuously in real-time \cite{021,020,037,035}. In \cite{021}, a smart wristband based stress detection model was proposed. In \cite{035}, a stress control system for the IoMT is proposed and in \cite{036} a smart-yoga pillow is studied to understand the relationship between stress and sleep. However, there may be very few research article that details technological solutions for detecting anxiety using low-cost consumer electronic components.

\begin{table*}[h]
\centering
\caption{Attributes of related work on Anxiety detection in older adults.}
\scalebox{1}{
\begin{tabular}{|c|c|c|c|c|c|}
\hline
& & & & &\\
\textbf{Work} & \textbf{Subjects} & \textbf{Acquisition} & \textbf{Signals} & \textbf{Target} & \textbf{Anxiety}\\
& & \textbf{Points} & & \textbf{Population} & \textbf{Reference} \\
\hline
Puli et al. \cite{022} & 15 & Chest and arm & ECG, ACC & Children and  & Physiological \\
& & & & Youth with ASD & Indicators \\
\hline
Wen et al. \cite{018} & 59 & Wrist and ankle & ECG & Younger Adults & Annotation by \\
& & & & & Audience Score\\
\hline
Zhang et al. \cite{019} & 92 & Head & EEG, eye- & Children and & Self-report \\
& & & movement & Adolescents & (SCARED)\\
\hline
Li et al. \cite{016} & 12 & Head & EEG & Younger & Predetermined \\
& & & & Adults & levels \\
\hline
Zheng  et al. \cite{015} & 20 & Head and nose & EEG, PPG & Younger Adults & STAI \\
\hline
McGinnis  et al. \cite{023} & 71 & NA & Audio & Children & Self-reported \\
& & &  & & and observed \\
\hline
\textbf{This Work} & \textbf{41} & \textbf{Wrist} & \textbf{Wrist-based signals} & \textbf{Older Adults} & \textbf{STAI}\\
& & &\textbf{(EDA and BVP)} & & \\
\hline 
\end{tabular} 
\label{related_work}
}
\end{table*} 

Table \ref{related_work} presents some of the existing works in anxiety detection along with a qualitative comparison with our proposed work. In \cite{022}, Puli et al. proposed a Kalman-like filter approach that fuses heart rate features extracted from ECG signal along with accelerometer signals for accurate anxiety level detection in the presence of motion. The target population for this work was children and youth who are clinically diagnosed with ASD (Autism Spectrum Disorder). The experiment was performed with 15 participants and the results and the analysis from the experiment showed that fusing accelerometer data with heart rate features was able to reduce false detection of anxiety to a significant level. Wen et al. \cite{018} proposed an anxiety detection model that uses features from the IBI series extracted from the ECG signal. The anxiety detection model proposed by the researchers uses a trained SVM model to classify samples as high or low anxiety states based on the extracted ECG features. In \cite{019}, Zhang et al. proposed fusing eye-movement features with EEG signals to increase the precision of anxiety detection. The researchers used group sparse canonical correlation analysis (GSCCA) in order to generate effective feature space containing both EEG and eye-movement features for improved performance of the anxiety detection model. The features were trained using SVM. Results and analysis showed that the fusing features from eye movement and EEG were able to improve the performance of the anxiety detection model. 

In \cite{016}, Li et al. have used extensive feature set from the EEG signal such as frequency domain, time domain, non-linear, and statistical features to train an SVM model to distinguish anxiety in 4 levels. The experiment was performed on 12 university students who were younger adults. A sensor fusion approach to classify anxiety in three levels was explored by Zheng et al. \cite{015} using EEG and PPG signals. 20 younger adults were used in this experiment during which the participants were required to complete two cycling-related tasks. Results from their analysis showed that features from both EEG and PPG performed better in detecting anxiety than when features from a single signal were used. In \cite{023}, McGinnis et al. explored the effectiveness of audio data in detecting anxiety among young children. Audio features were extracted from the audio signal recorded during a 3-minute speech task and the performance of several machine learning algorithms such as logistic regression, random forest, and SVM were evaluated.

\section{Proposed Method for Anxiety Detection}
\label{sec_proposed_method}
\begin{figure*}[h]
\centering
\includegraphics[trim= 0cm 0cm 0cm 0cm, scale=0.45]{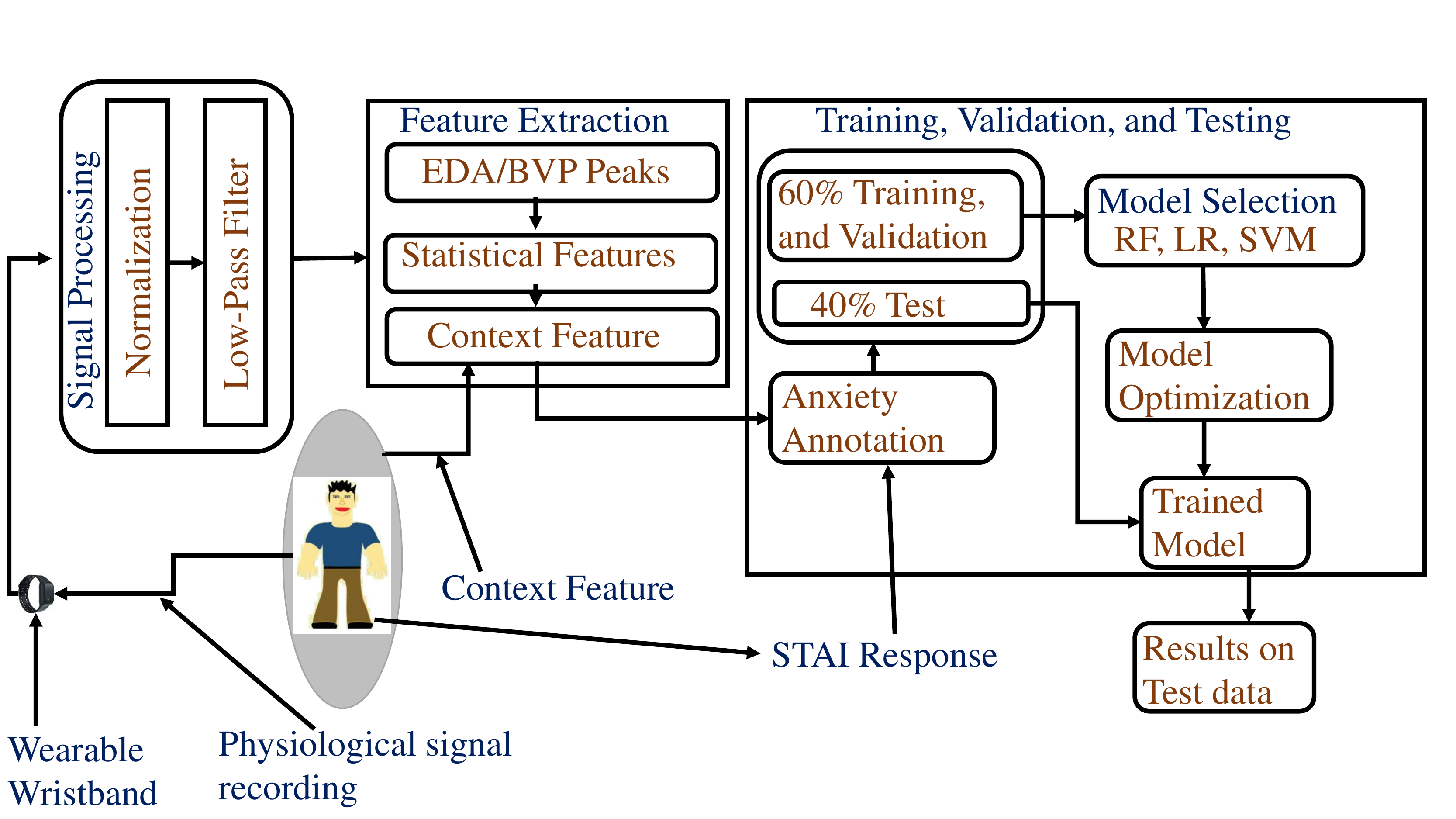}
\caption{Overview of the proposed method for anxiety detection. RF=Random Forest Classifier, LR=Logistic Regressiom, and SVM=Support Vector Machine. STAI=The State-Trait Anxiety Inventory}
\label{proposed_work_label}
\end{figure*}

This section will describe the proposed method  for anxiety detection. The proposed method for anxiety detection fuses an experimental context feature with features from a single physiological signal to train a machine learning classifier for accurately distinguishing between anxious and non-anxious states. The overview of the proposed method for anxiety detection is shown in Figure \ref{proposed_work_label}.

The major computational blocks of the proposed methods are: i) Signal Processing, (ii) Feature Extraction, and (iii) ML (Machine Learning) Modeling.

\subsection{Signal Processing}
This computational unit preprocesses the raw signal before the feature extraction stage. In this unit, the EDA and the BVP signals are first normalized by scaling the sensor values in the range [0,1]. After normalization, a low-pass Butterworth filter of order 5 was used to remove high-frequency components from the signal, which usually occurs as a result of motion artifact. For the EDA signal, the signal components that were higher than 1 Hz were cut off and for the BVP signal, the signal components that were higher than 10 Hz were cut off.  

\subsection{Feature Extraction}
In this section, we will describe the physiological features extracted from EDA, and BVP signals and the context feature used for training the machine learning model. Features are extracted from all the three phases of the experimental study. The detail about the phases of the experimental study are discussed in section \ref{exp_protocol_label}. For analysis, 3 minutes of data from pre-stress period, and about 15 minutes of data each from stress and relax phase were used. The physiological features are extracted from physiological signals using a running window of 30 seconds and an overlap of 15 seconds. On an average, 118 data points were generated from each participant and for 41 participants, a total of 4853 data points were obtained. The notation of the extracted physiological features from the three physiological signals are tabulated in Table \ref{feature_notation} 
\subsubsection{Feature Extraction from EDA Signal}
EDA signal measures the skin conductance of the skin which varies according to the sweating produced by the sweat glands. Arousal due to stressful situations results in the increased activity of the SNS (Sympathetic Nervous System) which leads to various physiological changes \cite{024}. Sweating is one of the physiological changes that is modulated by the SNS. Hence we hypothesize that using features from the EDA signal can be useful in detecting the onset of anxiety caused due to stressful events. Previous research has shown that the peak characteristic of the EDA signal is useful in detecting stress levels of an individual \cite{021,025}. Hence, we hypothesize that perceived anxiety due to emotional arousal resulting from stressful situations could also be modeled using the peak features from EDA. Peak detection algorithm \cite{026} is used to compute the amplitude, width, prominence of peaks occurring in a given window. Subsequently, statistical measures such as the mean, standard deviation, median, root mean square, minimum, and the maximum of peak amplitude, width, and prominence are computed to form the final set of 18 features. 
\subsubsection{Feature Extraction from BVP Signal}
The BVP signal is a measure of cardiovascular activity. BVP signal can be an indicator of cardiovascular arousal during exposure to stressful situations \cite{027}. Previous research that has used cardiovascular arousal as a measure to detect the onset of anxiety has resulted in satisfactory results. However, most of the existing research has used ECG to measure cardiovascular arousal in the context of anxiety detection. Since BVP-based features were successfully used previously in the context of stress detection \cite{021}, we hypothesize that BVP-based features can also be used to detect the onset of anxiety occurring as a result of stressful situations. Similar to EDA, peak detection algorithm \cite{026} is used to compute the width, amplitude, and prominence of the systolic peaks occurring in a given window. Along with statistical measures, the frequency of systolic peaks per minute is also computed resulting in a total of 17 features.    

\begin{table*}[h]
\centering
\caption{Notation and Description of the extracted features from EDA, and BVP.}
\scalebox{0.9}{
\begin{tabular}{|c|c|c|c|c|c|}

 \hline 
& & & & &\\
 \textbf{Signal} & \textbf{Feature} & \textbf{Feature} & \textbf{Signal} & \textbf{Feature} & \textbf{Feature} \\ 
 \textbf{Source} & \textbf{Notation} & \textbf{Description} & \textbf{Source} & \textbf{Notation} & \textbf{Description} \\ 
 & & & & &\\
 \hline 
 & & & & &\\
 EDA & $P_{amp}^{\bar{x}}$ & Mean of peak amplitudes & BVP & $S_{min}^{\# x}$ & No. of peaks per minute \\ 
 & & & & &\\
 EDA & $P_{amp}^{\tilde{x}}$ & Median of peak amplitudes & BVP & $S_{width}^{\bar x}$ & Mean of peak widths\\ 
 & & & & &\\
 EDA & $P_{amp}^{\sigma}$ & Standard deviation of peak amplitudes & BVP & $S_{width}^{\tilde x}$ & Median of peak widths \\ 
 & & & & &\\
 EDA & $P_{amp}^{\sqrt{\bar{x^{2}}}}$ & Root mean square of peak amplitudes & BVP & $S_{width}^{\sigma x}$ & Standard deviation of peak widths\\ 
 & & & & &\\
 EDA & $P_{amp}^{\vee x}$ & Maximum of peak amplitudes & BVP & $S_{width}^{\sqrt{\bar{x^{2}}}}$ & Root mean square of peak widths \\  
 & & & & &\\
 EDA & $P_{amp}^{\wedge x}$ & Minimum of peak amplitudes & BVP & $S_{width}^{\vee x}$ & Maximum of peak widths \\  
 & & & & &\\
 EDA & $P_{width}^{\bar{x}}$ & Mean of peak widths & BVP & $S_{width}^{\wedge x}$ & Minimum of peak widths \\ 
 & & & & &\\
 EDA & $P_{width}^{\tilde{x}}$ & Median of peak widths & BVP & $S_{prom}^{\bar x}$ & Mean of peak prominence\\  
 & & & & &\\
 EDA & $P_{width}^{\sigma}$ & Standard deviation of peak widths & BVP & $S_{prom}^{\tilde x}$ & Median of peak prominence \\ 
 & & & & &\\
 EDA & $P_{width}^{{\sqrt{\bar{x^{2}}}}}$ & Root mean square of peak widths & BVP & $S_{prom}^{\sigma x}$ & Standard deviation of peak prominence\\ 
 & & & & &\\
 EDA & $P_{width}^{\vee x}$ & Maximum of peak widths & BVP & $S_{prom}^{\sqrt{\bar{x^{2}}}}$ & Root mean square of peak prominence \\  
 & & & & &\\
 EDA & $P_{width}^{\wedge x}$ & Minimum of peak widths & BVP & $S_{prom}^{\vee x}$ & Maximum of peak prominence \\ 
 & & & & &\\
 EDA & $P_{prom}^{\bar{x}}$ & Mean of peak prominence & BVP & $S_{prom}^{\wedge x}$ & Minimum of peak prominence \\ 
 & & & & &\\
 EDA & $P_{prom}^{\tilde{x}}$ & Median of peak prominence & BVP & $S_{amp}^{\bar x}$ & Mean of peak amplitudes \\ 
 & & & & &\\
 EDA & $P_{prom}^{\sigma}$ & Standard deviation of peak prominence & BVP & $S_{amp}^{\sigma x}$ & Standard deviation of peak amplitudes\\  
 & & & & &\\
 EDA & $P_{prom}^{{\sqrt{\bar{x^{2}}}}}$ & Root mean square of peak prominence & BVP & $S_{amp}^{\sqrt{\bar{x^{2}}}}$ & Root mean square of peak amplitudes\\ 
 & & & & &\\
 EDA & $P_{prom}^{\vee x}$ & Maximum of peak prominence & BVP & $S_{amp}^{\vee x \Delta \wedge x}$ & Range of peak amplitudes \\ 
 & & & & &\\
 EDA & $P_{prom}^{\wedge x}$ & Minimum of peak prominence &   &  &  \\
 \hline 
 \end{tabular}  
\label{feature_notation}
}
\end{table*} 
\subsubsection{Context-Based Feature}
\label{context_feature_label}
The importance of context-awareness in the context of anxiety detection was studied in \cite{028}. The elicitation of anxious reaction due to stressful events is a subjective perception and is significantly affected by the context of the event \cite{034}. Hence, alongside features from physiological signals, using context-based features to train a machine learning model can result in better performance. For example, EDA measures emotional arousal and BVP measures cardiovascular arousal. However, arousals can be positive (because of positive stress or eustress), and negative (because of negative stress or distress). Hence, in situations where an individual is under positive arousal will not perceive the situation as an anxious situation. However, a machine learning model that models anxiety simply based on arousal might result in several false positive.

\begin{figure}[h]
\centering
\includegraphics[trim= 0cm 0cm 0cm 0cm, scale=0.23]{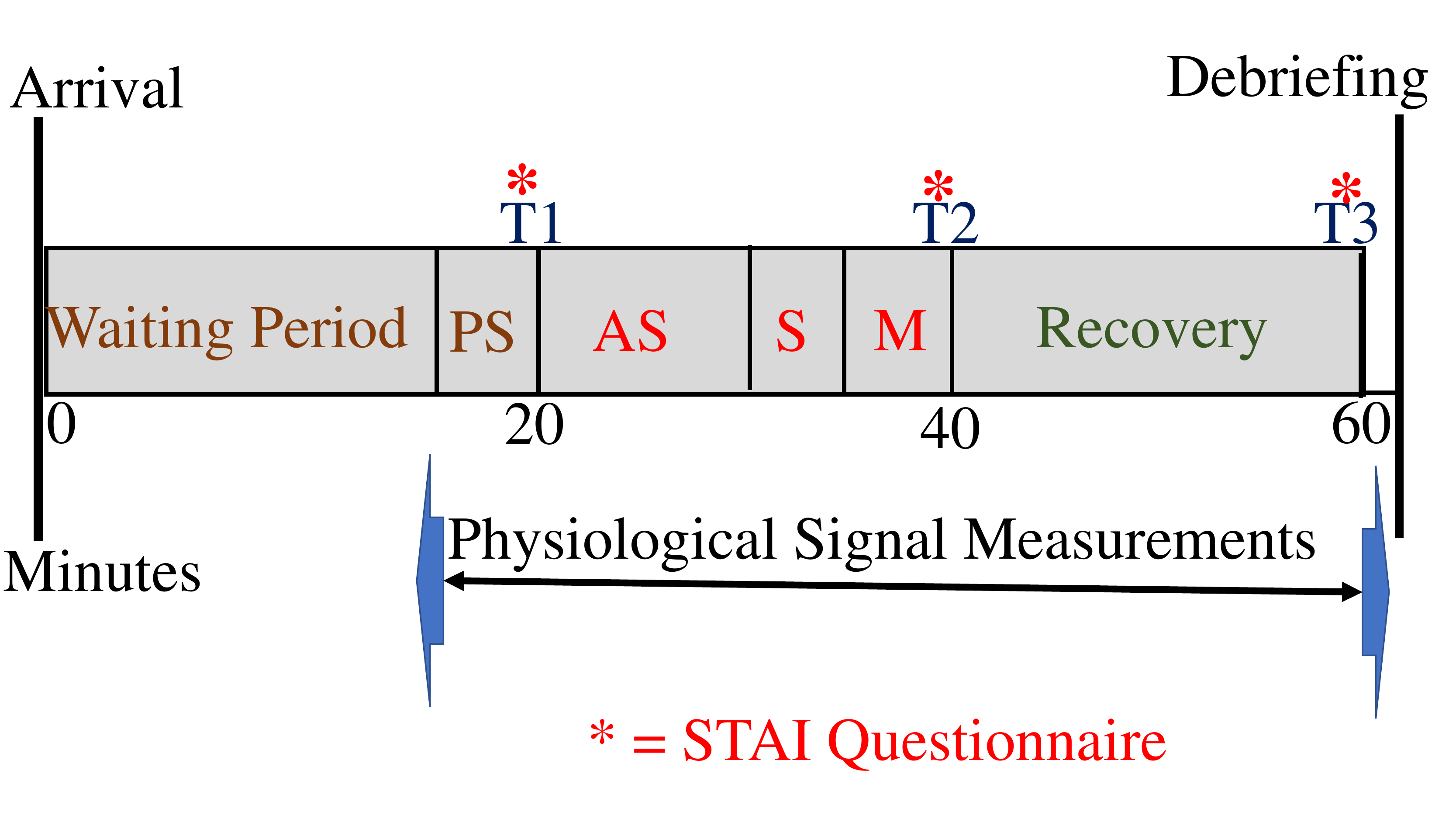}
\caption{Experimental Protocol }
\label{experimental_protocol_label}
\end{figure}

We hypothesize that encoding context information as a feature variable with the physiological features can help to distinguish the positive arousal from the negative arousal. To implement this, we encoded an experimental context feature with respective experimental segments of the physiological features. There are three distinct experimental context used in our study. The first experimental context is during the PS (Pre-Stress) period, during which the user is not subject to any stimulus (Figure \ref{experimental_protocol_label}). The second experimental context is during the AS (Anticipatory Stress), S (Stress), and M (Math), during which the user is subjected to stressful stimulus. The third context is during the Recovery period, during which the user is subjected to relaxation-based stimulus. The context feature used in our system is encoded as $-1$ for the experimental session when the user was not exposed to any kind of stimulus. The experimental session when the user is exposed to stressful stimulus is encoded as $0$. Finally, the session when the user is exposed to a relaxation-oriented stimulus is labeled as $1$. More details on the experimental protocol is described in Section \ref{exp_protocol_label}.

\subsection{Training, Validation, and Testing}
This section will describe the machine-learning-based modeling of anxiety. Primarily we will detail the training and testing protocol of the anxiety detection model adopted in this study. The three main components required for training and testing a machine learning model to distinguish between anxious and non-anxious states are the features from physiological signals, the experimental context feature, and the anxiety level ground truth. After the feature extraction, the feature samples are annotated with their respective anxiety states such as `Anxious' or `Not-Anxious' states. These anxiety states are determined from the STAI questionnaire filled by the user after each phase of the experimental protocol. Details on how anxiety states are determined from the STAI questionnaire will be described in detail in Section \ref{anxiety_ground_truth}.

For modeling the anxiety states with physiological features and the context feature, we have adopted a supervised machine learning approach. For model evaluation, the entire feature set is divided into a training set (50\%), a validation set (10\%), and a test set (40\%). The training and the validation set will be used to select the appropriate machine learning algorithm for anxiety states classification and hyperparameter tuning. The best performing machine learning algorithm in the training and validation phase will be chosen for testing on the test set.

\section{Experimental Setup}
\label{sec_experiment}
The objective of our experiment is to model the physiological changes resulting because of stressful situations with perceived anxiety levels for older adults. In this section, we will discuss the inclusion criteria for participants in our experiment, the data recording, ground truth estimation, and finally the experimental protocol adopted for the study. 

\subsection{Participants}
The participants recruited for this experiment were 41 older adults (26 females and 15 males) in the age range 60-80 with mean age $73.36 \pm 5.25$. Before recruiting for the experiment, the participants were screened for certain existing conditions which might interfere with the results of the experiment. These are heart-related conditions, post-traumatic disorder, anxiety disorders, unstable angina, Addison’s disease, or Cushing disease. Other than the existing conditions, participants were not recruited if they had any trouble
with daily activities, trouble following instructions, impaired consent
capacity, thinking problems, and have a diagnosis of dementia.
\subsection{Data Recording}
During the experiment, three different types of data were recorded: (i)physiological signals from wristband sensors, (ii) response to STAI questionnaire from the user, (iii) context-based feature from the experimental protocol. 
\subsubsection{Recording of Physiological Signals}
In our proposed anxiety detection, we have recorded two physiological signals during the experimental protocol. EDA signal is recorded with the help of two AgCl electrodes which maintains contact with the lower wrist of the user. The AgCl electrodes are attached to the strap of the wristband. The BVP signal is measured using the PPG sensor which uses a combination of red and green light for estimating the blood volume. The EDA and BVP signals were collected at a sampling rate of 4 Hz and 64 Hz respectively. 

The wristband device used in our study has a form factor of 110-190 mm and weighs around 25g. The wristband device executes the data transfer using low energy bluetooth (streaming mode) and has a flash memory (for recording mode) that can record upto 60 hours of data. In our experiment, the recording of EDA and BVP signals were conducted in recording mode to ensure collection of better signal quality for analysis. These signals were stored in in-device memory and were later synced to a desktop machine through the proprietary cloud interface after the end of each experiment. The collected physiological data did not contain any personal information about the participants.

\subsubsection{Recording of STAI Response}
\label{anxiety_ground_truth}
The STAI (The State-Trait Anxiety Inventory) is a commonly used self-feedback form to measure state and trait anxiety \cite{013}. The STAI is regarded as the gold standard for measuring anxiety \cite{030,031,029}. The participants were instructed to digitally respond to the 20 items STAI questionnaire intended to quantify state anxiety. Each of the participants was required to fill up the STAI questionnaire during the time points T1, T2, and T3 (Figure \ref{experimental_protocol_label}). The STAI questionnaire contains question fields such as ``I feel calm", ``I feel upset" etc and each of these fields is weighted on a scale of 1-4. The participants were asked to weigh each field according to how much they agree/disagree with the question fields. For example, a high level of agreement with a particular question field, ``I feel calm" will be weighted with 4 and a high level of disagreement for the same field will be weighted as 1. 

Scoring is done by summing all the scores of all fields. A higher score indicates a higher level of anxiety. Before scoring, the weights of all the question fields that are positive in nature, such as ``I feel calm", ``I feel relaxed" are reversed before they are summed. Hence for those fields, a weight of 4 will become 1, 3 becomes 2, and so on. The maximum score possible is 80 and the minimum score possible is 20. The mean and the standard deviation for all participants during the timestamps T1, T2, and T3 are shown in Table \ref{stai_stat}. 

\begin{table}[h]
\centering
\caption{Mean and Standard Deviation of the STAI response during each timestamps for the 41 participants}
\scalebox{1.2}{
\begin{tabular}{|c|c|c|}
\hline 
\textbf{TimeStamps} & \textbf{Mean} & \textbf{SD} \\ 
\hline 
T1 & 26.92 & 8.52 \\ 
\hline
T2 & 31.54 & 9.44 \\ 
\hline
T3 & 25.40 & 8.52 \\ 
\hline 

\end{tabular} 
\label{stai_stat}
}
\end{table}

To classify the score of a particular timestamp as anxious or non-anxious states, scores of all the participants are integrated and standardized. Subsequently, samples whose score is less than the population mean of the standardized score are classified as non-anxious states. Samples whose score is greater than the population mean are classified as anxious states. The statistics of the anxious state (``A") is $1.08 \pm 0.801$, and that of non-anxious state (``NA") is $-0.668 \pm 0.2144$. The p-value between the scores of anxious and non-anxious states is $5.29e^{-20}$ which indicates a significant statistical difference between the anxious and non-anxious states.

\subsubsection{Recording of Context Feature}
Since in this work we are evaluating a single contextual feature, the integration of context feature, based on the different phases of the experimental protocol (discussed in Section \ref{context_feature_label}) with physiological features is performed manually. The context feature used in this work is predetermined based on the experimental protocol and not on user feedback. In general, we recommend environmental context-based features that can be collected unobtrusively without requiring intervention from the user. 

\subsection{Experimental Protocol}
\label{exp_protocol_label}
To induce stress-related anxiety, the TSST (Trier Social Stress Test) is adopted because of its ability to induce stress naturally \cite{032}. During the experimental study, the participants were required to sit in a comfortable position and were instructed to minimize body movements as much as practicable during the study. The entire protocol begins with a waiting period, during which recording of demographic information, signing of the consent form, and briefing of experiment-related tasks. The phases of the experimental protocol after the waiting period can be divided into the pre-stress period, stress period, and recovery period. During the pre-stress period, baseline measurements are taken. The stress period begins with an AS (Anticipatory Stress) period where the participant is required to prepare a 5-minute speech task based on a given topic. The AS period lasts for 10 minutes, after which there is a subsequent speech and math task for 5 minutes each. The purpose of the stress stimulus is to increase the perceived feeling of anxiety in older adults naturally. The high mean value of the state scores after time point T2 (Table \ref{stai_stat}) shows that the experimental protocol was successful in inducing anxiety as a result of exposure to stressful situations. The final phase is the recovery phase in which the participants were exposed to a relaxing stimulus intended for the participants to return to the non-anxious state.

\section{Results and Analysis}
\label{sec_results}
This section will discuss and analyze the results of the proposed anxiety detection system. First, we will analyze the training and validation of the machine learning models in distinguishing between anxious and non-anxious states. We will then perform a qualitative analysis on the best-performing machine learning model in the training and validation phase. Subsequently, we will perform an extensive analysis of the selected model on its performance on the test data. Finally, we will analyze the feasibility of the trained model in classifying between anxious and non-anxious states in real-time. 

\begin{figure*}[h]
\centering
\includegraphics[trim= 0cm 0cm 0cm 0cm, scale=0.38]{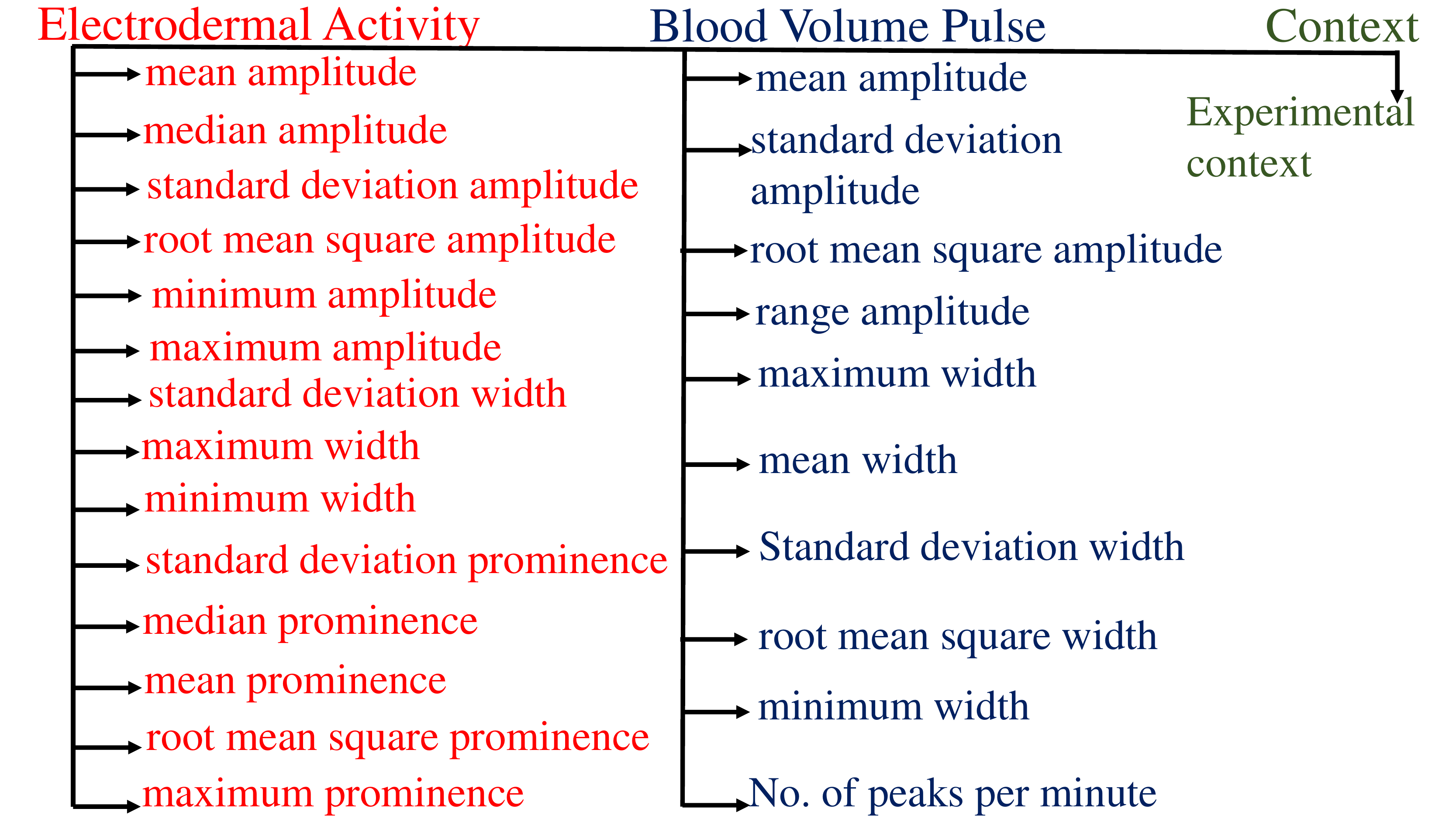}
\caption{Taxonomical representation of the selected features}
\label{selected_features_label}
\end{figure*}

\subsection{Training and Validation}
The training set, which is 50\% of the entire set is first used to select significantly correlated features with the anxiety labels obtained from the STAI score. This is done by taking Kendall’s tau correlation \cite{033} between each feature and the anxiety labels. The features which are only significantly correlated (p-value$<$0.05) are selected for training, validation, and testing the machine learning classifiers. Figure \ref{selected_features_label} shows the taxonomical representation of the selected features from each signal stream. 

From Figure \ref{selected_features_label}, we can see that 14 features out of the 18 features extracted from the EDA signal are found to be significantly correlated with the anxiety labels. The 14 features include the mean, median, RMS, maximum, and minimum of the EDA peak amplitudes along with standard deviation and maximum of EDA peak width and prominence. Other features selected from EDA are the minimum of EDA peak width, median, RMS, and mean of EDA prominence. From BVP signals, only the statistical measures of systolic peak width and amplitude were found to be significantly correlated with anxiety labels together with the frequency of systolic peaks per minute. The experimental context is also found to be significantly correlated with the anxiety labels. Hence, a total of 14 features were selected from EDA, and 10 features from BVP were selected for training the machine learning algorithms.

We have evaluated three machine learning algorithms in the training and validation phase. These machine learning algorithms are random forest classifier (RF), logistic regression (LR), and support vector machine (SVM). Table \ref{ml_perfm} shows the validation score of the three different machine learning algorithms on the validation set in distinguishing between anxious and non-anxious states. The hyperparameters of the machine learning models are tuned using an exhaustive grid search 5 fold cross-validation. 

\begin{table}[h]
\centering
\caption{Validation scores of different sensor combination with context-feature for different machine learning algorithms. RF=Random Forest, LR=Logistic Regression, SVM=Support Vector Machine, C=Context}
\scalebox{1.05}{
\begin{tabular}{|c|c|c|c|c|}
\hline 
\textbf{Algorithm} & \textbf{EDA} & \textbf{EDA$+$C} & \textbf{BVP} & \textbf{BVP$+$C} \\ 
\hline 
RF & 0.88 & 0.91 & 0.78 & 0.82 \\ 
\hline
LR & 0.51 & 0.61 & 0.53 & 0.61 \\ 
\hline
SVM & 0.37 & 0.58 & 0.45 & 0.61 \\ 
\hline 

\end{tabular} 
\label{ml_perfm}
}
\end{table}

\begin{figure}[h]
\centering
\includegraphics[trim= 0cm 0cm 0cm 0cm, scale=0.25]{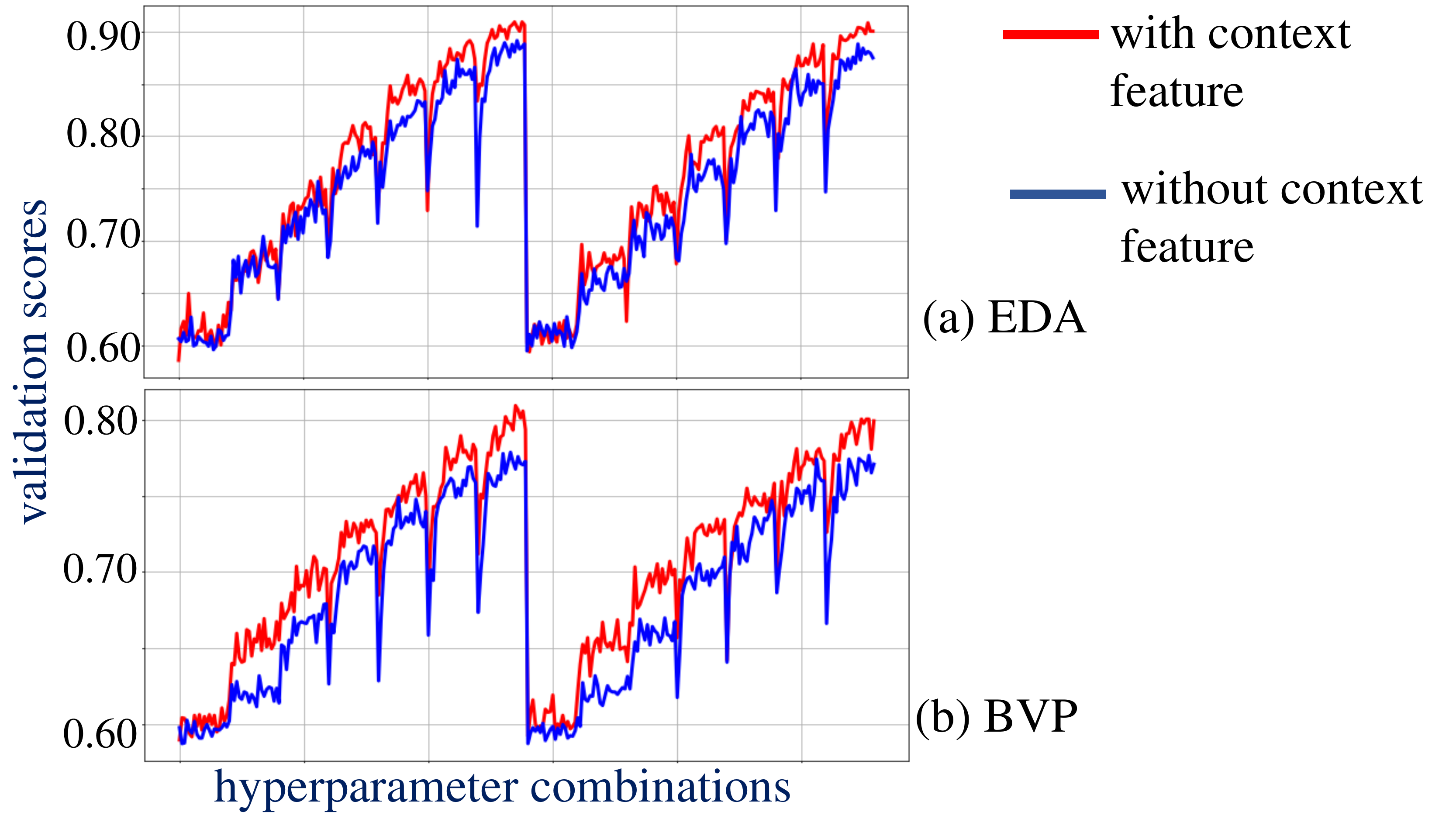}
\caption{Validation scores obtained by different hyperparameter combination for EDA, and BVP signals. The red lines indicate the scores obtained with combination of context feature and physiological features and the blue lines indicate the score obtained with only physiological features.}
\label{rf_valid_graph_label}
\end{figure}

From Table \ref{ml_perfm}, we can see that the RF classifier performed the best among LR and SVM in this case. RF algorithms achieved a validation score of 0.88 with EDA features. When the context feature is fused along with the EDA feature, the validation score increased to 0.91. Similar performance improvement is observed with the BVP signal when context features are fused with those features. Figure \ref{rf_valid_graph_label} shows the validation scores obtained by different hyperparameter combinations. From Figure \ref{rf_valid_graph_label}, we can visualize that for almost all hyperparameter combinations, the combination of context feature and physiological features performs better than using only the physiological features.  

Although LR and SVM did not perform as well in this dataset, the result is still encouraging because even for those classifiers, improvement in performance is observed when context feature is fused with the features of EDA and BVP signals. For example, LR achieved a 19\% increase in performance when the context feature is fused with EDA feature. Similar performance improvement is observed for the BVP signal as well. For SVM, the validation score increased by 56\% when the context feature is fused with the EDA feature. This highlights the importance of the context feature for distinguishing between anxious and non-anxious states. Since the RF model performed the best, we will be importing the optimized model for qualitative model evaluation and subsequently for testing and deployment.

\subsection{Evaluation of the Trained Random Forest Classifier}
In this section, we will perform a qualitative evaluation of the trained random forest model. Table \ref{model_hyperparameter} shows the hyperparameters of the 4 trained models using the 4 feature combinations. Table \ref{model_hyperparameter} shows the optimized model hyperparameters using cross-validation. 

\begin{table}[h]
\centering
\caption{Optimized hyperparameters for the six trained models using different feature combinations. C represents the context feature}
\scalebox{1.1}{
\begin{tabular}{|c|c|c|c|}
\hline 
\textbf{Trained} & \textbf{Split} & \textbf{Maximum} & \textbf{Number} \\ 
\textbf{Models} & \textbf{Criterion} & \textbf{Depth} & \textbf{Estimators} \\

\hline 
EDA & Gini & 7 & 20 \\ 
\hline 
EDA$+$C & Gini & 7 & 13 \\ 
\hline 
BVP & Gini & 7 & 18 \\ 
\hline 
BVP$+$C & Entropy & 7 & 12 \\ 
\hline 

\end{tabular} 
\label{model_hyperparameter}
}
\end{table}

\begin{table*}[h]
\centering
\caption{Performance metrics of the trained models on the test data}
\scalebox{1.2}{
\begin{tabular}{|c|c|c|c|c|c|}
\hline 

\textbf{Trained} & \textbf{Number of} & \textbf{F1-score} & \textbf{F1-score} & \textbf{Macro}  & \textbf{Accuracy}\\ 
\textbf{Models} & \textbf{Features} & \textbf{Anxious} & \textbf{Not-Anxious} & \textbf{F1-Score} & \textbf{(\%)} \\
\hline 
EDA & 14 &0.86 & 0.91 & 0.88 & 89\\ 
\hline 
\textbf{EDA$+$C} & \textbf{15} & \textbf{0.90} & \textbf{0.93} & \textbf{0.92} & \textbf{92}\\ 
\hline 
BVP & 10 & 0.71 & 0.83 & 0.77 & 78\\ 
\hline 
\textbf{BVP$+$C} & \textbf{11} & \textbf{0.77} & \textbf{0.86} & \textbf{0.82} & \textbf{83}\\ 
\hline 

\end{tabular} 
\label{test_perfm}
}
\end{table*}

The RF algorithm was optimized for the split criterion, maximum depth, and the number of estimators. The split criterion determines the quality of a split. Random forest partitions data into different nodes for decision-making by splitting based on feature values. The quality of the split is determined by either of the two criteria, Gini and entropy. Maximum depth is the maximum depth till which a decision tree is allowed to expand. The number of estimators refers to the number of decision trees in the forest.  

From Table \ref{model_hyperparameter}, we can observe that the optimized trained model that fuses context feature with physiological features have significantly less number of estimators than the ones that use only physiological features. For example, the trained model that uses EDA and the context feature has 35\% fewer estimators than the trained model that uses only EDA features. A similar decrease in the number of estimators is observed for the trained models that used BVP signal. The trained model that used the BVP and the context feature has only 12 estimators as compared to 18 estimators of the trained model that uses BVP features alone. This highlights the importance of the context feature in reducing the model complexity along with increasing the performance of the model. 

\subsection{Results on Test Data}
In this section, we will discuss the performance of the trained models (Table \ref{model_hyperparameter}). Table \ref{test_perfm} shows the results on the test data. The performance of the trained models are evaluated based on the models' capacity to distinguish between anxious (``A") and not-anxious (``NA") states. The performance metrics are the F1-score of the models' ability to detect the anxious states and not-anxious states, average macro F1-score of overall predictions, and overall accuracy. F1-score is the harmonic mean of precision and recall. Macro F1-score is an estimate of the overall performance estimate of the prediction model for both the classes. Macro average F1-score gives equal weight to both classes and hence this metric is not affected by the presence of class imbalance.   

\begin{figure*}[h]
\centering
\includegraphics[trim= 0cm 0cm 0cm 0cm, scale=0.45]{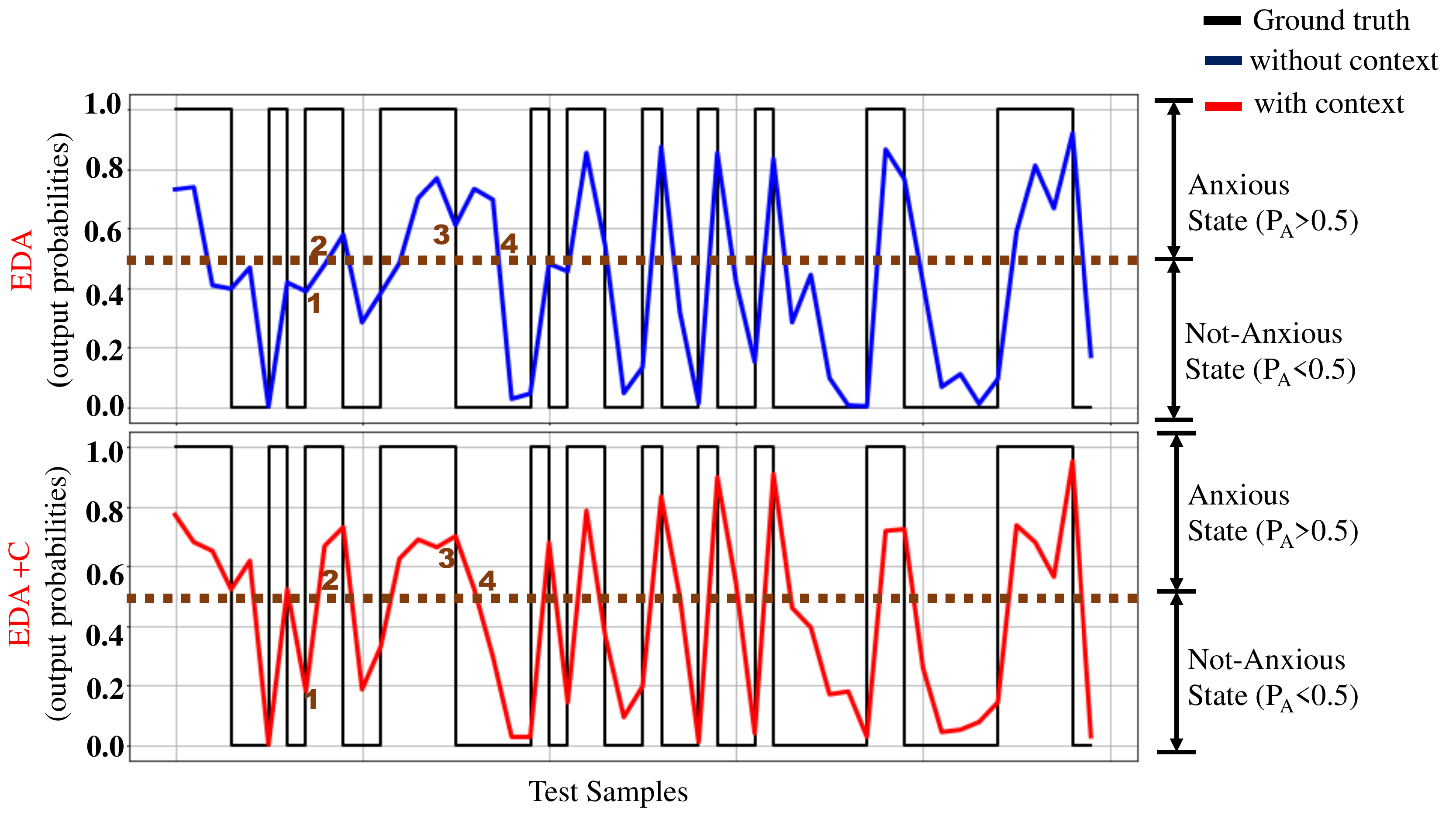}
\caption{Plot of the output probabilities of the machine learning models for EDA and EDA+C models}
\label{eda_w_wc_label}
\end{figure*}

\begin{figure*}[h]
\centering
\includegraphics[trim= 0cm 0cm 0cm 0cm, scale=0.45]{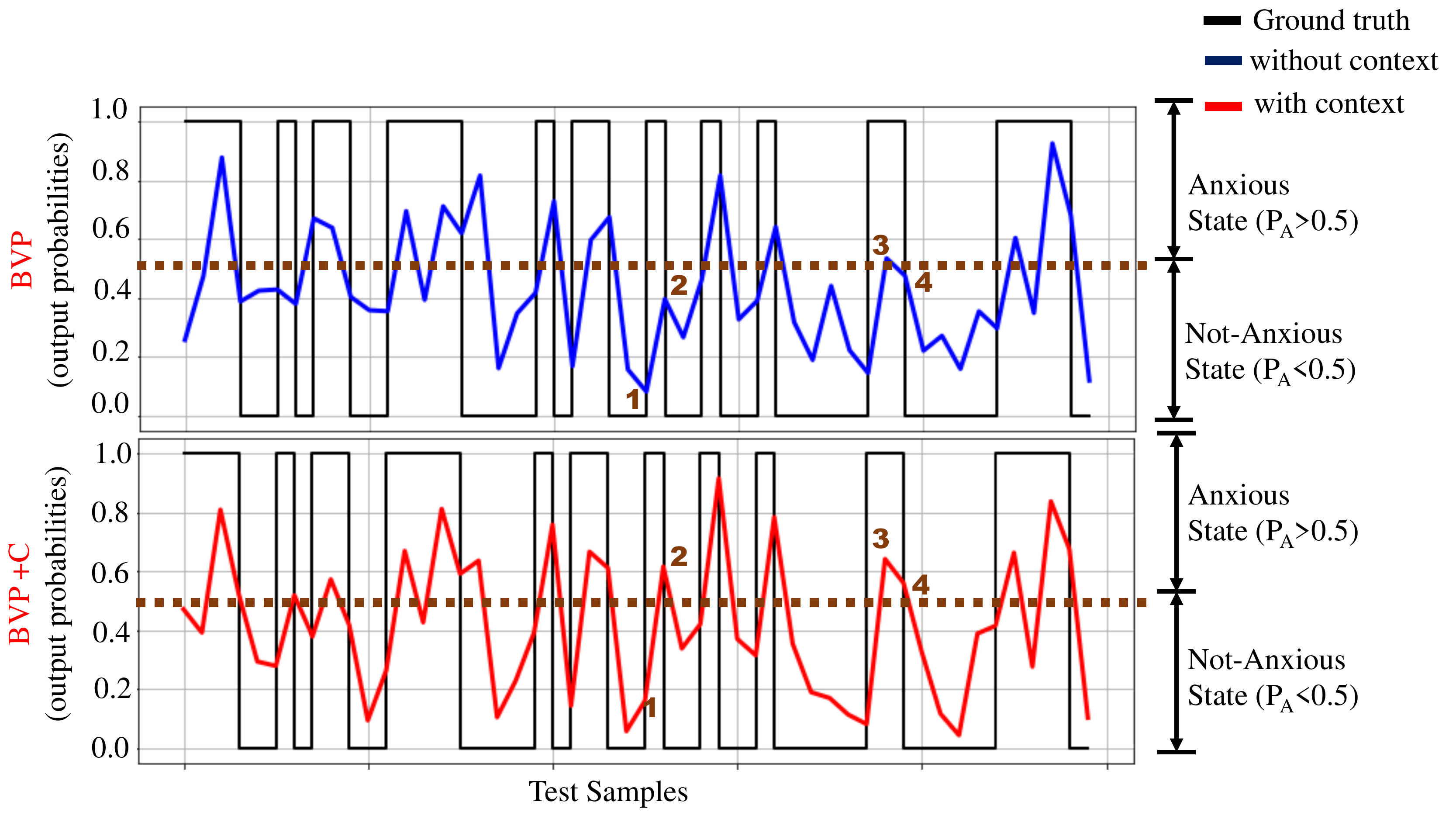}
\caption{Plot of the output probabilities of the machine learning models for BVP and BVP+C models}
\label{bvp_w_wc_label}
\end{figure*}

From Table \ref{test_perfm}, we can observe that both the hybrid physiological-context-based machine learning models-(EDA+C and BVP+C) performed better than the machine learning models that used only physiological signals (EDA and BVP). EDA+C model was able to increase the F1-score of the model in detecting the anxious state and non-anxious states by 4.65\% and 2.19\% respectively. The macro F1-score and overall accuracy of the EDA+C model were 4.54\% and 3.37\% higher than that of the EDA model. The BVP+C model achieved an 8.45\% and 3.61\% higher F1-score for the anxious and non-anxious states respectively than the BVP model. The macro F1-score and overall accuracy for the BVP+C model were also 6.49\% and 6.41\% higher than the BVP model respectively.

To further analyze how context-feature can improve the prediction of the machine learning models, we visualize the output probabilities of the machine learning models on the test set. The output probabilities for EDA and EDA+C models are shown in Figure \ref{eda_w_wc_label} and that for BVP and BVP+C are shown in Figure \ref{bvp_w_wc_label}. To classify a particular feature sample as either anxious or not-anxious, the random forest classifier machine learning models outputs two probabilities, $P_{A}$ and $P_{NA}$, where $P_{A}$ represents the probability of that feature sample to be classified as the anxious state (A), and $P_{NA}$ represents the probability of that feature sample to be classified as the not-anxious state. A threshold (usually 0.5) is set which acts as the decision boundary. Hence, for a particular sample, if $P_{A}>0.5$ then the model classifies that sample as anxious. In the same way, If $P_{A}<0.5$ implies $P_{NA}>0.5$, and hence, in this case, the feature sample will be classified as the not-anxious state. In other words, if the probability of the positive class (anxious class in this case) is closer to 1, the sample is classified as anxious and if the probability is closer to 0, the sample is classified as not-anxious. 

In Figure \ref{eda_w_wc_label}, the output probabilities of some of the test samples are plotted for EDA (top of the figure) and EDA+C (bottom of the figure). The decision boundary ($P$=0.5) is shown in the brown dotted line. The output of the trained models with context (plotted in red line), and without context (plotted in blue line) is shown along with the ground truth, the actual label (anxious or not-anxious). The ground truth is plotted in a black line as a step function. Similar plot is also obtained for BVP and BVP+C models in Figure \ref{bvp_w_wc_label}. 

From Figure \ref{eda_w_wc_label}, we can observe that, on addition of context feature, the output probability $P_{A}$ either decreases or increases to increase the confidence and hence avoid misclassification. For example, in Figure \ref{eda_w_wc_label}, for EDA model, we see that the test samples marked between 1 and 2 are mostly classified as ``NA" ($P_{A}<0.5$) even though the actual label is ``A". When the context feature is used along with the EDA features, the output probabilities of some of the test samples are increased above the threshold (0.5) and hence those samples are then correctly classified as ``A". Similarly, for the test samples in between the region 3 and 4, the actual label is ``NA". However the BVP model incorrectly classifies them as ``A" ($P_{A}>0.5$). When BVP+C model was used, the output probabilities of some of the test samples between the region 3 and 4 reduced below the threshold (0.5) and hence, those samples were correctly classified as ``NA". 

\begin{table*}
\centering
\caption{Simulation results on the size of the trained models and latency of the end-to-end processing for real-time anxiety detection}
\scalebox{1}{
\begin{tabular}{|c|c|c|c|c|c|c|}

\hline
 &  & \multicolumn{5}{c|}{ \textbf{Latency (seconds)}}\\
\cline{3-7}
\textbf{Trained} & \textbf{Model} &  \textbf{Data} &  \textbf{Signal} & \textbf{Feature} &   &  \textbf{Total}\\
 \textbf{Models}&  \textbf{Size(KB)} &  \textbf{Loading}  &  \textbf{Processing} &  \textbf{Extraction} & \textbf{Prediction}  & \textbf{Time}\\
\hline
EDA  & 136 &  0.004 & 0.003 & 0.002 & 0.009 & 0.014  \\
\hline
EDA+C  & 109 &  0.005 & 0.002 & 0.003 &  0.011 &  0.016  \\ 
\hline
BVP  & 196 &  0.037 & 0.002 & 0.002 &  0.013 &  0.05  \\ 
\hline
 BVP+C  & 115 &  0.037 & 0.003 & 0.002 &  0.007 &  0.05  \\ 
 \hline

  \hline
\end{tabular}
}
\label{latency_result}
\end{table*}

The same effect of the context-based feature can be visualized in Figure \ref{bvp_w_wc_label}. For example, all the test samples between the region 1 and 2 are classified as ``NA" ($P_{A}<0.5$)by the BVP model. However, when BVP+C model was used, the output probabilities of some of the test samples between the region and 1 and 2, are increased above the threshold (0.5) and hence are correctly classified as ``A". Similarly, for test samples between 3 and 4, the output probability of the BVP model is less than or equal to 0.5, even though the actual label is ``A". In this case, the classifier may not have enough confidence to classify samples as ``A" or might misclassify them as ``NA". When BVP+C model is used, the output probabilities increased above the threshold and hence, those samples were correctly classified as ``A". Based on this observations, it can be concluded that using context-based feature along with physiological features can perform better than that using only physiological feature. This reduces false positives and false negatives and provides a robust model for distinguishing between anxious and non-anxious states. 

\subsection{Real-Time Anxiety Detection}
To analyze the feasibility of the proposed method in detecting anxiety in real-time, the end-to-end process of anxiety detection is simulated for one minute. The simulation is performed on a computer with 16 GB RAM, and 3.3 GHz processor. Python 3.7 was used to simulate the results and the latency has been calculated using the time module of python. The end-to-end process consists of loading the data, signal processing, feature extraction, and prediction (Figure \ref{proposed_work_label}). The trained models (Table \ref{model_hyperparameter}) are imported as a .pkl file. Table \ref{latency_result} shows the simulation result for all the trained models (EDA, EDA+C, BVP, and BVP+C) in terms of latency, and the size of the trained model in memory.

Table \ref{latency_result} shows the memory of the trained models and the latency in seconds during each phase of the end-to-end processing. From Table \ref{latency_result}, we can observe that the size of all the trained models is just in the order of a few hundred KB (KiloByte). Further, the latency during each phase is less than a second. The context-based models also performed almost similar to non-context-based models in terms of latency while occupying less in-device memory.

\section{Conclusion}
\label{sec_conclusion}
In this paper, a hybrid physiological-context-based machine learning model for detecting anxiety in older adults is proposed. The proposed system uses two physiological sensors, EDA and PPG for recording EDA and BVP signals respectively during a TSST experimental protocol. Anxiety level ground truth was obtained from the STAI questionnaire Features from EDA and BVP signals were used to evaluate three machine learning models, random forest, logistic regression, and support vector machine in the training phase. The best performing random forest classifier was optimized in the training phase using an exhaustive grid search technique for evaluation on the test data. Results on the test data showed that the optimized context-based version of EDA and BVP models outperformed the non-context version of EDA and BVP models. The machine learning models that have used the context feature along with physiological features achieved higher F1-score for both the anxious and not-anxious states. Further, the feasibility of the proposed system for real-time anxiety detection in terms of latency and memory is discussed. 

In the existing literature, ECG and EEG signals are most commonly used for anxiety detection as these signals tend to be the most informative about anxiety levels. However, the acquisition setup required for recording these signals are usually complex and might be not be comfortable for daily use. Our proposed method for anxiety detection is based on simple wearable sensors such as EDA and PPG. These sensors can be easily integrated in low-cost consumer electronic devices such as a smart wristband. This will allow for continuous monitoring of anxiety levels with minimum obtrusiveness and will add to the comfort of the user. Further, our approach uses context feature in combination with physiological features instead of integrating features from other physiological signals. This is expected to minimize the design challenge faced when recording and processing information from multiple sensors in real-time. The current state of the proposed work on anxiety detection is suitable for controlled real-world settings where the activities (context) are more or less well defined. For example, a nursing home setting, or a training simulator could be some example use case for the proposed work. 

In the future, combination of more fine-grained context features could be studied along with the physiological features to better estimate the anxiety level of a person. This could be especially useful for detecting anxiety under different activity such as walking, running etc. Investigating ways to seamlessly incorporate different context features from environmental sensors or activity sensors with the physiological features for anxiety detection in uncontrolled real-world setting could be another future research direction. 

\section*{Acknowledgments}

This work was supported by the Kentucky Science and Engineering Foundation under Grant KSEF-3528-RDE-019.

\section{Compliance with Ethical Standards}
\textbf{Conflict of Interest}: The authors declare that they have no conflict of interest.\\
\textbf{Ethical approval}: All procedures performed in studies involving human participants were in accordance with the ethical standards of the institutional and/or national research committee and with the 1964 Helsinki declaration and its later amendments or comparable ethical standards.\\
\textbf{Informed consent}:Informed consent was obtained from all individual participants included in the study.

\bibliographystyle{spmpsci} 
\bibliography{references}

\begin{thebibliography}{10}
\providecommand{\url}[1]{{#1}}
\providecommand{\urlprefix}{URL }
\expandafter\ifx\csname urlstyle\endcsname\relax
  \providecommand{\doi}[1]{DOI~\discretionary{}{}{}#1}\else
  \providecommand{\doi}{DOI~\discretionary{}{}{}\begingroup
  \urlstyle{rm}\Url}\fi

\bibitem{034}
Adolph, D., Meister, L., Pause, B.M.: {{C}ontext counts! social anxiety
  modulates the processing of fearful faces in the context of chemosensory
  anxiety signals}.
\newblock Front Hum Neurosci \textbf{7}, 283 (2013)

\bibitem{032}
Birkett, M.A.: The trier social stress test protocol for inducing psychological
  stress.
\newblock Journal of visualized experiments: JoVE (56) (2011)

\bibitem{030}
Dalal, K.S., Chellam, S., Toal, P.: Anaesthesia information booklet: Is it
  better than a pre-operative visit?
\newblock Indian journal of anaesthesia \textbf{59}(8), 511 (2015)

\bibitem{001}
Daviu, N., Bruchas, M.R., Moghaddam, B., Sandi, C., Beyeler, A.:
  Neurobiological links between stress and anxiety.
\newblock Neurobiology of Stress \textbf{11}, 100191 (2019).
\newblock \doi{https://doi.org/10.1016/j.ynstr.2019.100191}

\bibitem{008}
Edition, F., et~al.: Diagnostic and statistical manual of mental disorders.
\newblock Am Psychiatric Assoc \textbf{21} (2013)

\bibitem{004}
Emdin, C.A., Odutayo, A., Wong, C.X., Tran, J., Hsiao, A.J., Hunn, B.H.:
  {{M}eta-{A}nalysis of {A}nxiety as a {R}isk {F}actor for {C}ardiovascular
  {D}isease}.
\newblock Am J Cardiol \textbf{118}(4), 511--519 (2016)

\bibitem{003}
Fawcett, J.: {{T}he detection and consequences of anxiety in clinical
  depression}.
\newblock J Clin Psychiatry \textbf{58 Suppl 8}, 35--40 (1997)

\bibitem{024}
Fechir, M., Schlereth, T., Purat, T., Kritzmann, S., Geber, C., Eberle, T.,
  Gamer, M., Birklein, F.: {{P}atterns of sympathetic responses induced by
  different stress tasks}.
\newblock Open Neurol J \textbf{2}, 25--31 (2008)

\bibitem{007}
Flint, A.J.: {{A}nxiety and its disorders in late life: moving the field
  forward}.
\newblock Am J Geriatr Psychiatry \textbf{13}(1), 3--6 (2005)

\bibitem{010}
van Hout, H.P., Beekman, A.T., de~Beurs, E., Comijs, H., van Marwijk, H.,
  de~Haan, M., van Tilburg, W., Deeg, D.J.: {{A}nxiety and the risk of death in
  older men and women}.
\newblock Br J Psychiatry \textbf{185}, 399--404 (2004)

\bibitem{031}
Kayikcioglu, O., Bilgin, S., Seymenoglu, G., Deveci, A.: State and trait
  anxiety scores of patients receiving intravitreal injections.
\newblock Biomedicine hub \textbf{2}(2), 1--5 (2017)

\bibitem{033}
Kendall, M.G.: A new measure of rank correlation.
\newblock Biometrika \textbf{30}(1/2), 81--93 (1938)

\bibitem{012}
Klainin-Yobas, P., Oo, W.N., Suzanne~Yew, P.Y., Lau, Y.: {{E}ffects of
  relaxation interventions on depression and anxiety among older adults: a
  systematic review}.
\newblock Aging Ment Health \textbf{19}(12), 1043--1055 (2015)

\bibitem{027}
Kulkarni, S., O'Farrell, I., Erasi, M., Kochar, M.S.: {{S}tress and
  hypertension}.
\newblock WMJ \textbf{97}(11), 34--38 (1998)

\bibitem{016}
{Li}, Z., {Wu}, X., {Xu}, X., {Wang}, H., {Guo}, Z., {Zhan}, Z., {Yao}, L.: The
  recognition of multiple anxiety levels based on electroencephalograph.
\newblock IEEE Transactions on Affective Computing pp. 1--1 (2019).
\newblock \doi{10.1109/TAFFC.2019.2936198}

\bibitem{017}
{Liu}, H., {Wen}, W., {Zhang}, J., {Liu}, G., {Yang}, Z.: Autonomic nervous
  pattern of motion interference in real-time anxiety detection.
\newblock IEEE Access \textbf{6}, 69763--69768 (2018).
\newblock \doi{10.1109/ACCESS.2018.2880465}

\bibitem{023}
{McGinnis}, E.W., {Anderau}, S.P., {Hruschak}, J., {Gurchiek}, R.D.,
  {Lopez-Duran}, N.L., {Fitzgerald}, K., {Rosenblum}, K.L., {Muzik}, M.,
  {McGinnis}, R.S.: Giving voice to vulnerable children: Machine learning
  analysis of speech detects anxiety and depression in early childhood.
\newblock IEEE Journal of Biomedical and Health Informatics \textbf{23}(6),
  2294--2301 (2019).
\newblock \doi{10.1109/JBHI.2019.2913590}

\bibitem{009}
Mehta, K.M., Simonsick, E.M., Penninx, B.W., Schulz, R., Rubin, S.M.,
  Satterfield, S., Yaffe, K.: {{P}revalence and correlates of anxiety symptoms
  in well-functioning older adults: findings from the health aging and body
  composition study}.
\newblock J Am Geriatr Soc \textbf{51}(4), 499--504 (2003)

\bibitem{021}
{Nath}, R.K., {Thapliyal}, H.: Smart wristband-based stress detection framework
  for older adults with cortisol as stress biomarker.
\newblock IEEE Transactions on Consumer Electronics \textbf{67}(1), 30--39
  (2021).
\newblock \doi{10.1109/TCE.2021.3057806}

\bibitem{037}
Nath, R.K., Thapliyal, H., Caban-Holt, A.: Machine learning based stress
  monitoring in older adults using wearable sensors and cortisol as stress
  biomarker.
\newblock Journal of Signal Processing Systems pp. 1--13 (2021)

\bibitem{029}
Nigussie, S., Belachew, T., Wolancho, W.: Predictors of preoperative anxiety
  among surgical patients in jimma university specialized teaching hospital,
  south western ethiopia.
\newblock BMC surgery \textbf{14}(1), 1--10 (2014)

\bibitem{028}
{Panagiotakopoulos}, T.C., {Lyras}, D.P., {Livaditis}, M., {Sgarbas}, K.N.,
  {Anastassopoulos}, G.C., {Lymberopoulos}, D.K.: A contextual data mining
  approach toward assisting the treatment of anxiety disorders.
\newblock IEEE Transactions on Information Technology in Biomedicine
  \textbf{14}(3), 567--581 (2010).
\newblock \doi{10.1109/TITB.2009.2038905}

\bibitem{006}
Papadimitriou, G.N., Kerkhofs, M., Kempenaers, C., Mendlewicz, J.: {{E}{E}{G}
  sleep studies in patients with generalized anxiety disorder}.
\newblock Psychiatry Res \textbf{26}(2), 183--190 (1988)

\bibitem{026}
Pedregosa, F., Varoquaux, G., Gramfort, A., Michel, V., Thirion, B., Grisel,
  O., Blondel, M., Prettenhofer, P., Weiss, R., Dubourg, V., et~al.:
  Scikit-learn: Machine learning in python.
\newblock the Journal of machine Learning research \textbf{12}, 2825--2830
  (2011)

\bibitem{022}
{Puli}, A., {Kushki}, A.: Toward automatic anxiety detection in autism: A
  real-time algorithm for detecting physiological arousal in the presence of
  motion.
\newblock IEEE Transactions on Biomedical Engineering \textbf{67}(3), 646--657
  (2020).
\newblock \doi{10.1109/TBME.2019.2919273}

\bibitem{036}
Rachakonda, L., Bapatla, A.K., Mohanty, S.P., Kougianos, E.: Sayopillow:
  Blockchain-integrated privacy-assured iomt framework for stress management
  considering sleeping habits.
\newblock IEEE Transactions on Consumer Electronics \textbf{67}(1), 20--29
  (2021).
\newblock \doi{10.1109/TCE.2020.3043683}

\bibitem{035}
Rachakonda, L., Mohanty, S.P., Kougianos, E.: ifeliz: An approach to control
  stress in the midst of the global pandemic and beyond for smart cities using
  the iomt.
\newblock In: 2020 IEEE International Smart Cities Conference (ISC2), pp. 1--7
  (2020).
\newblock \doi{10.1109/ISC251055.2020.9239028}

\bibitem{020}
{Rachakonda}, L., {Mohanty}, S.P., {Kougianos}, E., {Sundaravadivel}, P.:
  Stress-lysis: A dnn-integrated edge device for stress level detection in the
  iomt.
\newblock IEEE Transactions on Consumer Electronics \textbf{65}(4), 474--483
  (2019).
\newblock \doi{10.1109/TCE.2019.2940472}

\bibitem{025}
{Setz}, C., {Arnrich}, B., {Schumm}, J., {La Marca}, R., {Tröster}, G.,
  {Ehlert}, U.: Discriminating stress from cognitive load using a wearable eda
  device.
\newblock IEEE Transactions on Information Technology in Biomedicine
  \textbf{14}(2), 410--417 (2010).
\newblock \doi{10.1109/TITB.2009.2036164}

\bibitem{013}
Spielberger, C.D.: State-trait anxiety inventory.
\newblock The Corsini encyclopedia of psychology pp. 1--1 (2010)

\bibitem{002}
Takagi, Y., Sakai, Y., Abe, Y., Nishida, S., Harrison, B.J.,
  Martínez-Zalacaín, I., Soriano-Mas, C., Narumoto, J., Tanaka, S.C.: A
  common brain network among state, trait, and pathological anxiety from
  whole-brain functional connectivity.
\newblock NeuroImage \textbf{172}, 506--516 (2018).
\newblock \doi{https://doi.org/10.1016/j.neuroimage.2018.01.080}

\bibitem{005}
Vytal, K., Cornwell, B., Arkin, N., Letkiewicz, A., Grillon, C.: The complex
  interaction between anxiety and cognition: insight from spatial and verbal
  working memory.
\newblock Frontiers in Human Neuroscience \textbf{7}, 93 (2013).
\newblock \doi{10.3389/fnhum.2013.00093}

\bibitem{018}
{Wen}, W., {Liu}, G., {Mao}, Z.H., {Huang}, W., {Zhang}, X., {Hu}, H., {Yang},
  J., {Jia}, W.: Toward constructing a real-time social anxiety evaluation
  system: Exploring effective heart rate features.
\newblock IEEE Transactions on Affective Computing \textbf{11}(1), 100--110
  (2020).
\newblock \doi{10.1109/TAFFC.2018.2792000}

\bibitem{011}
Wetherell, J.L., Thorp, S.R., Patterson, T.L., Golshan, S., Jeste, D.V., Gatz,
  M.: {{Q}uality of life in geriatric generalized anxiety disorder: a
  preliminary investigation}.
\newblock J Psychiatr Res \textbf{38}(3), 305--312 (2004)

\bibitem{019}
{Zhang}, X., {Pan}, J., {Shen}, J., {Din}, Z.U., {Li}, J., {Lu}, D., {Wu}, M.,
  {Hu}, B.: Fusing of electroencephalogram and eye movement with group sparse
  canonical correlation analysis for anxiety detection.
\newblock IEEE Transactions on Affective Computing pp. 1--1 (2020).
\newblock \doi{10.1109/TAFFC.2020.2981440}

\bibitem{015}
{Zheng}, Y., {Wong}, T.C.H., {Leung}, B.H.K., {Poon}, C.C.Y.: Unobtrusive and
  multimodal wearable sensing to quantify anxiety.
\newblock IEEE Sensors Journal \textbf{16}(10), 3689--3696 (2016).
\newblock \doi{10.1109/JSEN.2016.2539383}

\bibitem{014}
Zung, W.W., Gianturco, J.A.: {{P}ersonality dimension and the {S}elf-{R}ating
  {D}epression {S}cale}.
\newblock J Clin Psychol \textbf{27}(2), 247--248 (1971)

\end{thebibliography}

\end{document}